\documentclass[pra,twocolumn,titlepage,nofootinbib,amsmath
,preprintnumbers]{revtex4}%
\usepackage{amsmath}
\usepackage{amsfonts}
\usepackage{amssymb}
\usepackage{graphicx}%
\usepackage{color}

\begin{document}

\def\lbar{\lambda\hskip-4.5pt\vrule height4.6pt depth-4.3pt width4pt}

\title{Light-by-light-scattering contributions to the Lamb shift in light muonic atoms}

\author{Evgeny Yu. Korzinin}
\author{Valery A. Shelyuto}
\affiliation{D.~I. Mendeleev Institute for Metrology, St.Petersburg,
190005, Russia}
\affiliation{Pulkovo Observatory, St.Petersburg, 196140, Russia}
\author{Vladimir G. Ivanov} \affiliation{Pulkovo
Observatory, St.Petersburg, 196140, Russia}
\author{Robert Szafron}
\affiliation{Technische Universit\"at M\"unchen, Fakult\"at f\"ur Physik, 85748 Garching, Germany}
\author{Savely~G.~Karshenboim}
\email{savely.karshenboim@mpq.mpg.de}
\affiliation{Ludwig-Maximilians-Universit{\"a}t, Fakult{\"a}t f\"ur Physik, 80799 M\"unchen, Germany}
\affiliation{Max-Planck-Institut f\"ur Quantenoptik, Garching, 85748, Germany}
\affiliation{Pulkovo Observatory, St.Petersburg, 196140, Russia}


\today
\preprint{TUM-HEP-1174/18}

\begin{abstract}
We consider one-loop light-by-light-scattering contributions to the Lamb
shift of the $1s, 2s, 2p$ states in light muonic hydrogen like atoms at $Z\leq10$.
The contributions are of the order $\alpha^5m_\mu$ (with diverse dependence
on the nuclear charge $Z$). Those include the contributions of the so-called
Wichmann-Kroll potential ($\alpha(Z\alpha)^4m_\mu$), the virtual Delbr\"uck
scattering ($\alpha^2(Z\alpha)^3m_\mu$), etc. The results are obtained in a
nonrelativistic approximation. For the calculation of the
virtual-Delbr\"uck-scattering contribution, we have constructed an effective
potential in the coordinate space which may be applied to other calculations
in muonic atoms.
\end{abstract}
\maketitle

\section{Introduction\label{s:introduction}}

Muonic atoms give an opportunity to develop and test a bound-state QED theory and probe a nuclear structure with a specific range of parameters not available with ordinary [electronic] atoms.
Recently the accuracy of the measurement of the $2s-2p$ Lamb shift in some
light hydrogen like muonic atoms has been dramatically improved~\cite{science:h,science:d}. The QED theory of the energy levels in muonic atoms is somewhat
different from that in ordinary atoms.
The Bohr radius in muonic atoms is comparable with the Compton wave length of an electron.
Because of that, an important role is played by the diagrams with the closed electron loops.
Those contributions are specific
for muonic atoms. The most important are those due to vacuum polarization.
Their contribution to the energy is of the order $\alpha(Z\alpha)^2m$.

\begin{figure}[htbp]
\begin{center}
\includegraphics[height=3cm]{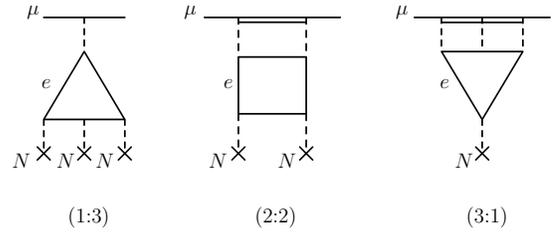}
\end{center}
\caption{Characteristic diagrams induced by the light-by-light scattering.
The double horizontal line is for the nonrelativistic Coulomb Green's function
of a muon. \label{fig:LbL}}
\end{figure}

Effects of the virtual light-by-light scattering contribute to higher orders.
There are three types of such contributions, characteristic diagrams
which are presented in Fig.~\ref{fig:LbL}. They are all of the order
$\alpha^5m$, but their dependence on the value of the nuclear $Z$ charge
is different.

The $\alpha(Z\alpha)^4m$ contribution (see the graph 1:3 in Fig.~\ref{fig:LbL}) is the so-called Wichmann-Kroll (WK) contribution, which has been studied for
a while (see, e.g.,~\cite{bor_rin,VASH-book}). A number of the results have
been achieved for muonic atoms using certain numerical approximations of
the exact WK potential. In particular, the approximations, introduced
in~\cite{huang} and \cite{bor_rin} on the basis of the results of
numerical integration
in \cite{vogel}, were numerously applied (e.g., in \cite{pach,bor_rin,EGS}).
The result for the $2p-2s$ Lamb shift with the accuracy sufficient for
applications in $\mu$H was found in \cite{EGS} and confirmed in \cite{bor_h,LbL1,LbL2}. In \cite{LbL1,LbL2} the result was also confirmed by direct
calculations. The WK contributions to the $n=2$ Lamb shift for some other
light muonic atoms are obtained in, e.g., \cite{bor_d,bor_he,VP2rel}.

The $\alpha^2(Z\alpha)^3m$ term is due to the virtual Delbr\"uck scattering (see the 2:2 diagram in Fig.~\ref{fig:LbL}). It has also been studied for quite a long period (see, e.g., \cite{bor_rin,VASH-book}). Still, some questions have
been resolved only recently \cite{LbL1}.

The initial calculations were based on a so-called scattering approximation~\cite{scattering} (where the Coulomb muon propagator is substituted for a free one). The substitution by itself is incorrect (see, e.g., discussion in \cite{VASH-book,LbL1}); however, the formulas which were eventually used in the numerical calculations were nevertheless correct (see below). Results on the contribution to the Lamb shift in some light atoms were published, e.g., in \cite{bor_rin,bor_h}, but they were not very accurate.

The third type of contributions (see the 3:1 plot in Fig.~\ref{fig:LbL}) have not been calculated until recently. It was studied in \cite{LbL1,LbL2}, where also the virtual-Delbr\"uck-scattering contribution was found with a sufficient accuracy for several light muonic atoms.

A kind of theorem on the 2:2 and 3:1 contributions was announced in \cite{LbL2} and proven in \cite{LbL1}. The papers considered an approximation of a static muon, where its nonrelativistic propagator is presented with a $\delta$ function over the energy. It was proven that the approximation is a valid one. We discuss the accuracy of the approximation in this paper (see Sec.~\ref{s:potential}). Using that approximation \cite{LbL2,LbL1}, the results on the 2:2 and 3:1 contributions to the Lamb shift in muonic hydrogen, deuterium and helium ions have
been found (see \cite{VP2rel} for $\mu$T). It was also demonstrated that
the related limit can be achieved both from the diagrams with the bound-muon Green's function (as shown in Fig.~\ref{fig:LbL}) and from those with the free
Green's function (as were used in the scattering approximation in \cite{bor_rin,bor_h}). As far as the static-muon approximation is applicable, one may use
both types of diagrams with the same result, which validates the working formulas used in \cite{bor_rin,bor_h}.

In this paper we consider the effective potential for the virtual-Delbr\"uck-scattering contribution to the Lamb shift in light muonic two-body atoms. We use the representation of the potential in momentum space in terms of an integral over Feynman parameters~\cite{LbL1} and study the effective potential in the
coordinate space by means of an analytic Fourier transform and subsequent
numerical integrations over the Feynman parameters. For the effective
potential in the coordinate space, we find both asymptotics (at $r\ll 1/m_e$ and $r\gg 1/m_e$). (Here and throughout the paper we apply the relativistic units in which $\hbar=c=1$.) Eventually, we fit the numerical results and asymptotics,
obtained here. The approximation is accurate at the level of $10^{-3}$ in the
area where the muon wave function of low states is localized.

Our main results are related to the virtual-Delbr\"uck-scattering contribution to the Lamb shift; however, we present numerical results for all three light-by-light  (LbL)
contributions (see Fig.~\ref{fig:LbL}), because their comparison can be useful.

The $2p-2s$ Lamb-shift interval cannot be successfully measured in all the
two-body muonic atoms (because of the range of the interval); however,
the theory of the Lyman-$\alpha$ transition is very similar.
The data on such gross-structure transitions play an important role in
determination of the rms charge radius of a large variety of elements
(see, e.g., \cite{radii}).
In this paper we tabulate the virtual light-by-light-scattering contribution
to the Lamb shift of the $1s, 2s, 2p$ states which is sufficient for the
calculation of both the $2p-2s$ interval and the energy of the $2p-1s$
transition.
The considered range of the nuclear charge is $Z=1,...10$.

\section{The effective potential and the static-muon approximation \label{s:potential}}

As demonstrated in \cite{LbL1}, once we can neglect various contributions to the muon propagator, such as the binding energy and those related to
momentum transfer [between the muon and the electron loop] in comparison with its energy transfer $q_0$, we arrive
at the nonrelativistic propagator reduced to $\delta(q_0)$. For
$Z\alpha m_\mu/n\leq m_e$ ($n$ is the principle quantum number), the energy
transfer is determined by the $m_e$ scale. In the opposite case, when
$Z\alpha m_\mu/n\geq m_e$, the characteristic value of $q_0$ is determined by
the value of the momentum (in the LbL loop), which in its turn is determined
by the characteristic atomic momentum $Z\alpha m_\mu$. That means that once
$Z\alpha\ll1$, we can apply the static-muon approximation.
(In \cite{LbL1} we considered a stronger condition $(Z\alpha)^2m_\mu\ll m_e$.)
All that is related, indeed, to only 2:2 and 3:1 contributions. The standard
WK contribution does not require any conditions on the muon but only on
the static regime of the nucleus. Those conditions are weaker and the
validity of the WK potential is due to relativistic-recoil effects, i.e.,
due to corrections which are of higher order in both small parameters of
the two-body Coulomb problem, $Z\alpha$ and $m_\mu/M$, where $M$ is
the nuclear mass.

\begin{figure}[thbp]
\vspace{-4pt}
\begin{center}
\resizebox{0.99\columnwidth}{!}{\includegraphics{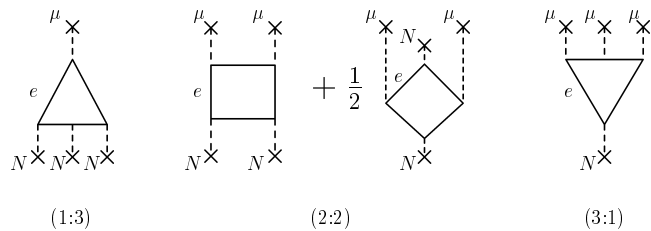}}
\end{center}
\vspace{-7pt} \caption{"Double-external-field" approximation with a static
nucleus and a static muon.}
\label{fig:stat}       
\end{figure}

Once the static-muon approximation is applicable, we arrive at a
"double-external-field" limit, the diagrams for which are presented
in Fig.~\ref{fig:stat}.
In particular, that allows us to immediately set a relation between
the 3:1 contribution and the 1:3 one (WK);
\begin{equation}\label{eq:31:13}
\Delta E_{3:1}(ns)=\frac1{Z^2}\,\Delta E_{1:3}(ns)\;,
\end{equation}
since the related integrands differ by their normalization only.
Note that Eq. (\ref{eq:31:13}) is correct only under the static-muon
approximation. The corrections beyond the approximation are of different
orders for $\Delta E_{3:1}$ and $\Delta E_{1:3}$.

The potential for the 1:3 contribution was studied for a while and there are
a number of efficient approximations, such as those mentioned above
from~\cite{huang} and \cite{bor_rin}. (Still, we revisit the problem in
Sec.~\ref{s:num}.)

An effective potential for the 2:2 contribution, an evaluation of which is
the main purpose of this paper, is considered in detail in the next section.

\section{The effective potential for the virtual-Delbr\"uck-scattering
contribution\label{s:asy}}

Following \cite{LbL1}, the contribution of virtual Delbr\"uck scattering
to the Lamb shift in light muonic atoms can be presented in terms of a
certain potential. In the momentum space the result reads \cite{LbL1}
\begin{equation}
\Delta E_{2:2} =  \int \frac{d^3{\bf q}}{(2\pi)^3}\,V_{2:2}({\bf q}^2)\,F({\bf q}^2)
\end{equation}
where the potential $V({\bf q}^2)$ is discussed in details in \cite{LbL2}
and
 \begin{eqnarray}
 F_{nl}({\bf q}^2)&=&\int \frac{d^3{\bf p}}{(2\pi)^3}\,\left(\Psi_{nl}({\bf p}-{\bf q})\right)^*\Psi_{nl}({\bf p})\nonumber\\
 &=&\int d^3{\bf r}\,
 \left(\Psi_{nl}({\bf r})\right)^*\,e^{-i({\bf q}\cdot{\bf r})}\Psi_{nl}({\bf
 r})
  \end{eqnarray}
is the form factor of the atomic $nl$ state, while $\Psi_{nl}({\bf p})$
is its nonrelativistic Coulomb wave function (with the reduced mass $m_r$).

\onecolumngrid
~\\
The potential $V_{2:2}({\bf q}^2)$ is presented in momentum space as an integral over the Feynman parameters \cite{LbL1}
\begin{eqnarray}\label{eq:v22:q}
V_{2:2}({\bf q}^2)&=& \frac{3}{4\pi}\;\alpha^2(Z\alpha)^2\,\int_0^1dx\int_0^1dy\int_0^1dz
  \int_0^1du\int_0^1dv\int_0^1dw\int_0^1dt\nonumber\\
&&\times\sum_{k=1,2}
   \left\{
 \frac{{\cal B}^{(k)}_{2:2}}{\left(s_{2:2}^{(k)}\,{\bf q}^2+m_e^2\right)} +
\frac{{\cal C}^{(k)}_{2:2}\,{\bf q}^2}
 {\left(s^{(k)}_{2:2}\,{\bf q}^2+m_e^2\right)^2}
   \right.
+   \left. \frac{{\cal D}^{(k)}_{2:2}\,{\bf q}^4}
 {\left(s_{2:2}^{(k)}\,{\bf q}^2+m_e^2\right)^3}
 \right\}\;,
\end{eqnarray}
where ${\cal B}^{(k)}_{2:2}$, ${\cal C}^{(k)}_{2:2}$, ${\cal
D}^{(k)}_{2:2}$, and $s_{2:2}^{(k)}$ are bulky dimensionless functions
of those parameters considered in \cite{LbL1}. The parameter $k$ is to
distinguish two diagrams contributing to $V_{2:2}$: $k=1$ stands for the left 2:2 graph (see Fig.~\ref{fig:stat}) and $k=2$ is for the right one.

\begin{figure}[htbp]
\begin{center}
\resizebox{0.95\columnwidth}{!}{\includegraphics[clip]{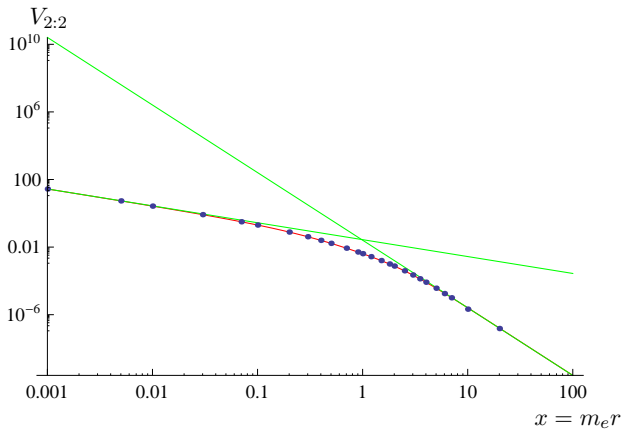}}
\end{center}
\caption{The "data" (i.e., the results of our numerical calculation of (\ref{eq:v22:r}) in coordinate space), their asymptotics, and the fit from (\ref{eq:fit}) (see below). The potential $V_{2:2}$ is given in units of $-\alpha^2(Z\alpha)^2m_e$, and the distance is characterized with $x=r m_e$.\label{fig:fit}}
\end{figure}

The dependence on ${\bf q}^2$ is simple, which allows us to immediately perform
the Fourier transformation
\begin{equation}\label{q->x}
  V_{2:2}(r)=\frac{4\pi}{r}\int_0^\infty  \frac{dq}{(2\pi)^3}\,q\,\sin(qr)V_{2:2}({\bf q}^2)
 \end{equation}
and to obtain a result in the coordinate space, which reads
\begin{eqnarray}\label{eq:v22:r}
V_{2:2}(r)&=& \frac{3}{4\pi}\,\alpha^2(Z\alpha)^2\,\int_0^1dx\int_0^1dy\int_0^1dz
  \int_0^1du\int_0^1dv\int_0^1dw\int_0^1dt\sum_{k=1,2}\exp\left(-\frac{m_er}{\sqrt{s_{2:2}^{(k)}}}\right)\nonumber\\
&&\times
   \left\{
 \frac{ {\cal B}^{(k)}_{2:2}}{4\pi s_{2:2}^{(k)}\,r}
+\frac{{\cal C}^{(k)}_{2:2}}{\left(s_{2:2}^{(k)}\right)^3}\,\frac{2s_{2:2}^{(k)}-m_er\sqrt{s_{2:2}^{(k)}}}
{8\pi\,r }
 + \frac{{\cal D}^{(k)}_{2:2}}{\left(s_{2:2}^{(k)}\right)^4}\, \frac{8s_{2:2}^{(k)}-m_er\left(7\sqrt{s_{2:2}^{(k)}}-m_e r\right)}
{32\pi\,r }
 \right\}
\;.
\end{eqnarray}
The explicit representation of the potential $V_{2:2}(r)$ is cumbersome and
for practical applications we further look for an efficient approximate formula.
To derive it we first find the value of the potential in certain points in
the coordinate space (see Fig.~\ref{fig:fit}) and then fit them with
a Pad\'e approximation.~\\
\twocolumngrid

To improve the accuracy of the fit, prior to fitting, we look for the asymptotics. The potential behaves as $\propto r^{-1}$ at short distances, as one should
expect from (\ref{eq:v22:r}), while at long distances it is $\propto r^{-4}$. The general situation is illustrated in the plot in Fig.~\ref{fig:fit}. The range of characteristic values of $x$, which are of interest for light muonic atoms, is summarized in Table~\ref{t:x:char}.

\begin{table}[htbp]
\begin{tabular}{c|cccc}
\hline
Ion&Z&$\kappa$&$x_1$&$x_2$\\ \hline
 $^{1}${H}   & 1 & 1.356  & 0.737  & 2.950 \\
 $^{2}${H}   & 1 & 1.428  & 0.700  & 2.800 \\
 $^{3}${H}   & 1 & 1.454  & 0.688  & 2.751 \\
 $^{3}${He}  & 2 & 2.908  & 0.344  & 1.375 \\
 $^{4}${He}  & 2 & 2.935  & 0.341  & 1.363 \\
 $^{6}${Li}  & 3 & 4.443  & 0.225  & 0.900 \\
 $^{7}${Li}  & 3 & 4.455  & 0.224  & 0.898 \\
 $^{9}${Be}  & 4 & 5.960  & 0.1678 & 0.671 \\
 $^{10}${B}  & 5 & 7.460  & 0.1341 & 0.536 \\
 $^{11}${B}  & 5 & 7.467  & 0.1339 & 0.536 \\
 $^{12}${C}  & 6 & 8.968  & 0.1115 & 0.446 \\
 $^{13}${C}  & 6 & 8.975  & 0.1114 & 0.446 \\
 $^{14}${N}  & 7 & 10.48  & 0.0954 & 0.382 \\
 $^{15}${N}  & 7 & 10.48  & 0.0954 & 0.382 \\
 $^{16}${O}  & 8 & 11.99  & 0.0834 & 0.334 \\
 $^{17}${O}  & 8 & 11.99  & 0.0834 & 0.334 \\
 $^{18}${O}  & 8 & 12.00  & 0.0834 & 0.333 \\
 $^{19}${F}  & 9 & 13.50  & 0.0741 & 0.296 \\
 $^{20}${Ne} & 10 & 15.00 & 0.0667 & 0.267 \\
 $^{21}${Ne} & 10 & 15.01 & 0.0666 & 0.267 \\
 $^{22}${Ne} & 10 & 15.01 & 0.0666 & 0.266 \\
 \hline
\end{tabular}
  \caption{Characteristic differences of the wave functions of the low states ($1s, 2s, 2p$) in light two-body muonic atoms. Here, $\kappa=Z\alpha m_r/m_e$ is the characteristic momentum of the muonic states in the units of $m_e$, while $x_n=n^2/\kappa$ is the characteristic radius of the $nl$ state in units of $\lbar_e=\hbar/m_ec$.}\label{t:x:char}
\end{table}

The short-distance asymptotic coefficient can be directly established from
(\ref{eq:v22:r}) in a rather straightforward way. The result of the numerical
integration reads
\begin{equation}\label{eq:low}
V_{2:2}(r\ll1/m_e)\simeq -0.027\,565(13)\,\frac{\alpha^2(Z\alpha)^2}{r}\;.
\end{equation}

The large-distance asymptotic behavior is not that simple to establish
from (\ref{eq:v22:r}). Considering the LbL contributions (see Fig.~\ref{fig:LbL}) in the $t$ channel, we note that some pure photonic intermediate states are
possible there, which sets the branch point for $t=-{\bf q}^2$ to zero and eventually leads to a certain $r^{-p}$ behavior at large distances for each of the LbL potentials (cf. \cite{WK1}). In the case of $V_{2:2}(r)$ in thedd form of
(\ref{eq:v22:r}), that technically means a singularity of the effective
dispersion-relation variable (cf. (\ref{eq:v22:q})) at $m_e^2/s^{(k)}_{2:2}=0$,
which should transform the exponential factor in (\ref{eq:v22:r}) to $r^{-p}$.

Fortunately, the asymptotic behavior of the 2:2 potential can be successfully studied in a different way; namely, we find it from the virtual-Delbr\"uck-scattering amplitude for soft photons \cite{rev:vD1,rev:vD2,vDs} (cf. \cite{LbL:CS}) as
\begin{eqnarray}\label{eq:high}
V_{2:2}(r\gg 1/m_e)&\simeq& -\frac{59}{2304}\,\frac{\alpha^2(Z\alpha)^2 m_e}{(m_er)^4}\nonumber\\
&\simeq& -0.025\,61\,
\frac{\alpha^2(Z\alpha)^2 m_e}{(m_er)^4}\;.
\end{eqnarray}

\onecolumngrid
~\\
With the asymptotic coefficients in hand, we fit the numerical results. The fit reads
\begin{equation}
\label{eq:fit}
V_{2:2}^{\rm approx}(r)=-\frac{\alpha^2(Z\alpha)^2}{r}
\frac{7.236+0.3099 x+2.561 x^2}{262.5+902.0 x+751.7 x^2+458.6
x^3+2.62 x^4+100x^5} \;,
\end{equation}
where $x= m_er$. The fit has $\chi^2=9.5$ for 22 degrees of freedom.
We estimate the accuracy of the fit as $1\times 10^{-3}$ for $x\leq 1$. In the interval of $1<x<10$ the uncertainty gradually increases to a few percent level. For higher $x$, thanks to the correct asymptotic behavior, the error does not exceed that level.
~\\
\twocolumngrid

As an independent test of our fit, we compare the results obtained by using
the fit for the $n=2$ Lamb shift in the lightest two-body muonic atoms with the direct ones \cite{LbL1,LbL2} (see Table~\ref{t:fit:valid:2:2}). The results are in perfect agreement within our estimation of the uncertainty of the fit as
$10^{-3}$.

\begin{table}
 \begin{tabular}{lc|rr}
 \hline
 Atom, state & $x$ & \multicolumn{2}{c}{contribution [meV]} \\[1ex]
 \cline{3-4}
       & & \multicolumn{1}{c}{Eq. (\ref{eq:fit})}  & \multicolumn{1}{c}{direct} \\[1ex]
   \hline
   $\mu$H ($2s$)       &    2.95        & $ 0.001\,791(4)$ & $0.001\,793(3)$    \\[1ex]
   $\mu$H ($2p$)       &                & $ 0.000\,642(1)$ & $0.000\,642(2)$  \\[1ex]
   $\mu$D ($2s$)       &    2.80        & $-0.001\,966(4)$    &  $-0.001\,968(3)$ \\[1ex]
   $\mu$D ($2p$)       &                & $-0.000\,733(1)$    &  $-0.000\,734(2)$ \\[1ex]
   $\mu^4$He$^+$ ($2s$)    &    1.36    & $ 0.027\,28(3)$ & $0.027\,31(4)$   \\[1ex]
   $\mu^4$He$^+$ ($2p$)    &            & $ 0.015\,88(2)$ &  $0.015\,88(3)$  \\[1ex]
   \hline
 \end{tabular}
  \caption{The 2:2 contributions to the $2s$ and $2p$ Lamb shift in light muonic atoms. The results of direct calculations are taken from \cite{LbL1,LbL2}. The uncertainty of the integration over the fit in (\ref{eq:fit}) is the statistical one. The error due to the static-muon approximation is the same for the direct calculations and for those from the fit. The characteristic value of $x$
is $x=x_2$.}\label{t:fit:valid:2:2}
\end{table}

The virtual-Delbr\"uck-scattering situation is very different from the WK one.
As mentioned, the WK potential $V_{1:3}(r)$ \cite{WK1} is valid when one can
neglect the recoil effects, i.e., it is a result of an expansion not only in
$Z\alpha$, but also in $m/M$. Because of the recoil nature of the corrections,
the WK potential is applicable in both ordinary and muonic atoms. In the
former we are interested in a large range of distances at $x\gg 1$, while the latter deals only with $x\sim 1$ or $x\ll 1$. The 2:2 potential is applicable only for muonic atoms \cite{LbL1,LbL2} and therefore the area with $x\gg 1$ and even with $x\geq 1$ is of low interest. It still may appear in evaluation of
the energy for the highly excited states with $n^2/Z\gg1$, but most of the
applications rely on a study of the lower states with $n=1,2$. For such
states the accuracy of the Pad\'e approximation (\ref{eq:fit}) is at the
level of $10^{-3}$. Note, that this is the accuracy of the approximation of $V_{2:2}(r)$ potential. Meanwhile, the very applicability of that potential due to the
static muon approximation has lower accuracy (see above).

As an example of applicability of the $x\gg1$ area to practical cases,
we mention neutral antiprotonic helium, where the characteristic size of
the antiproton orbit is comparable with the $1s$ orbit of an electron in
a hydrogen atom (see, e.g., \cite{antih}).

\section{Numerical results\label{s:num}}

The purpose of the paper is a derivation of an effective potential for the
2:2 contribution to the muonic-atom Lamb shift at medium $Z$, which has
been done in the previous section. It is interesting to compare the numerical
results with those from other LbL terms, and in particular, with the WK ones.

There are two fits for the  WK potential for the muonic atoms,
which are available in literature. (The potential is valid by itself for
ordinary and muonic atoms; however, the purpose of the fit determines
the range of the distances of interest (see above).) One of them is \cite{huang}
\begin{eqnarray}\label{eq:wk:app}
V_{1:3}(r)&=&0.3617\,\frac{\alpha(Z\alpha)^2}{\pi}\frac{Z\alpha}{r} \,  \exp\biggl[0.3728\, x \nonumber\\
&&-\sqrt{2.906+11.4\,x+4.417\, x^2}\biggr].
\end{eqnarray}

Another fit applied in numerical calculations in muonic atoms is \cite{bor_rin}
\begin{equation}\label{eq:fit:r}
V_{1:3}=\frac{\alpha(Z\alpha)^3}{\pi^3 r} \left\{
\begin{array}{ll}
\frac{-0.1755 + 0.1559 x + 0.0880 x^2}{x^6}  & \mbox{for}~~x\geq 1  \\
\frac{0.649 - 0.208 x}{1.374 x^3+1.41 x^2+2.672 x+1} & \mbox{for}~~x\leq 1  \\
\end{array}
\right.\;.
\end{equation}

Both fits are based on numerical calculations by Vogel \cite{vogel} for
the interval of $0.1<x\leq 1$ and in that area the fits well agree with
the numerical results (at the level of $10^{-3}$). They both utilize the
known leading asymptotic term at low $x$. They are different in area $x>1$.
The advantage of (\ref{eq:wk:app}) is more smooth behavior around $x=1$
and therefore a better extrapolation to the low end of the $x>1$ interval,
while the fit in (\ref{eq:fit:r}) accommodates the asymptotic term at
$x\gg 1$ and is better at high end of the interval.

\onecolumngrid
~\\
We use our own fit of Vogel's data \cite{vogel}
\begin{equation}\label{eq:fit:wk}
V_{\rm app.\,WK}(x)=\frac{\alpha(Z\alpha)^3}{r}\frac{5.026 + 0.02676 x
+ 0.2829 x^2}{240.0 + 725.4 x +
 542.2 x^2 + 649.8 x^3 + 150.2 x^4 + 9.457 x^5 + 100 x^6}\,,
\end{equation}
which fits the data for $0.1<x\leq 1$ with a fractional uncertainty better than $10^{-3}$ and correctly reproduces the asymptotics at low $r$ \cite{WK1} (see also \cite{blomqvist,bell}) and at high $r$ \cite{WK1} (see also \cite{huang,manakov}). In contrast to the fit (\ref{eq:fit:r}) from \cite{bor_rin},
our fit in (\ref{eq:fit:wk}) has smooth behavior at 6taround $x=1$.
~\\
\twocolumngrid

The application of the fits to the $n=2$ Lamb shift in  muonic hydrogen is rather questionable (see Table~\ref{t:x:char}), since we essentially need to
integrate over an interval outside of the data area of \cite{vogel}, which
was used to derive the fit. The smooth behavior at around $x=1$ and a correct
$x\gg 1$ asymptotics (mentioned above) should deliver a reasonable result,
but its accuracy is unclear.

Previously, while calculating the results for muonic hydrogen, deuterium,
and helium \cite{LbL1,LbL2,VP2rel} we have used a direct calculation instead
of the fits. To verify the accuracy of the previous fits and our fit,
we compare our results of a direct calculation and the results from the fits
for $2s, 2p$ for a few light atoms where the characteristic values of $x$ are the largest (see Table~\ref{t:fit:valid}).  The error of our fit is about 1\%,
while for the others it is at a few-percent level. Eventually we estimate
the accuracy of our fit as follows; at $0.1<x\leq1$ it is below $1\times 10^{-3}$, and it gradually reduces for $x<0.1$ and $x>1$ down to a 1\% level.

\begin{table*}
 \begin{tabular}{lc|r|r|r|r}
   \hline
 Atom, state & $x$ & \multicolumn{4}{c}{contribution [meV]}  \\[1ex]
\cline{3-6}
       & & \multicolumn{1}{c|}{Eq. (\ref{eq:wk:app})}
       & \multicolumn{1}{c|}{Eq. (\ref{eq:fit:r})}
       & \multicolumn{1}{c|}{Eq. (\ref{eq:fit:wk})}
       & \multicolumn{1}{c}{direct} \\[1ex]
   \hline
      $\mu$H ($2s$)       &    2.95        & 0.001\,240 & 0.001\,238 & 0.001\,243 & $0.001\,2472(7)$    \\[1ex]
   $\mu$H ($2p$)       &                & 0.000\,2196 & 0.000\,2196 & 0.000\,2270 & $0.000\,228\,87(4)$  \\[1ex]
   $\mu$D ($2s$)       &    2.80        & 0.001\,362 & 0.001\,358 & 0.001\,364 & $0.001\,3693(7)$   \\[1ex]
   $\mu$D ($2p$)       &                & 0.000\,2609 & 0.000\,2609 & 0.000\,2691 & $0.000\,271\,23(4)$  \\[1ex]
   $\mu$He$^4$ ($2s$)    &    1.36        & 0.037\,67 & 0.037\,30 & 0.037\,69 & $0.037\,833(22)$   \\[1ex]
   $\mu$He$^4$ ($2p$)    &                & 0.017\,68 & 0.017\,68 & 0.017\,82 & $0.017\,8676(15)$  \\[1ex]
   \hline
 \end{tabular}
  \caption{The WK contributions to the $2s$ and $2p$ Lamb shift in light muonic atoms. The results of direct calculations are taken from \cite{LbL1,LbL2}. The uncertainty of the fits for $x>1$ is {\em a priori\/} unclear and not shown.
  \label{t:fit:valid}}
\end{table*}

The results for $n=1,2$ states in a two-body muonic atom are summarized
in Tables~\ref{t:1s}, \ref{t:2s}, and \ref{t:2p} for all three LbL
contributions (the 1:3, 2:2, 3:1 ones). The uncertainty of the fits is
discussed above, as well as the uncertainty of the static-muon approximation.

\begin{table*}
\begin{tabular}{cc|ccccc}
 \hline
 Ion&Z&$\Delta E_{1:3}$&$\Delta E_{2:2}$&$\Delta E_{3:1}$&$\Delta E_{\rm LbL}(1s)$&$\Delta E_{\rm LbL}(1s)$ \\
     &  & [$\alpha^3(Z\alpha)^2m_r$] & [$\alpha^3(Z\alpha)^2m_r$] & [$\alpha^3(Z\alpha)^2m_r$] & [$\alpha^3(Z\alpha)^2m_r$] &[meV]\\
\hline
 $^{1}${H} & 1 & 0.005\,804 & $-0.008\,095$ & 0.005\,804 & 0.003\,513 & 0.006\,903 \\
 $^{2}${H} & 1 & 0.006\,073 & $-0.008\,410$ & 0.006\,073 & 0.003\,736 & 0.007\,734 \\
 $^{3}${H} & 1 & 0.006\,167 & $-0.008\,520$ & 0.006\,167 & 0.003\,814 & 0.008\,038 \\
 $^{3}${He} & 2 & 0.040\,28 & $-0.026\,23$ & 0.010\,07 & 0.024\,13 & 0.2034 \\
 $^{4}${He} & 2 & 0.040\,49 & $-0.026\,35$ & 0.010\,12 & 0.024\,26 & 0.2063 \\
 $^{6}${Li} & 3 & 0.1118 & $-0.047\,97$ & 0.012\,43 & 0.076\,30 & 1.474 \\
 $^{7}${Li} & 3 & 0.1120 & $-0.048\,02$ & 0.012\,44 & 0.076\,39 & 1.479 \\
 $^{9}${Be} & 4 & 0.2227 & $-0.071\,50$ & 0.013\,92 & 0.1651 & 5.704 \\
 $^{10}${B} & 5 & 0.3737 & $-0.096\,03$ & 0.014\,95 & 0.2926 & 15.81 \\
 $^{11}${B} & 5 & 0.3738 & $-0.096\,06$ & 0.014\,95 & 0.2927 & 15.83 \\
 $^{12}${C} & 6 & 0.5656 & $-0.1213$ & 0.015\,71 & 0.4600 & 35.87 \\
 $^{13}${C} & 6 & 0.5657 & $-0.1213$ & 0.015\,72 & 0.4601 & 35.90 \\
 $^{14}${N} & 7 & 0.7988 & $-0.1471$ & 0.016\,30 & 0.6681 & 71.00 \\
 $^{15}${N} & 7 & 0.7989 & $-0.1471$ & 0.016\,30 & 0.6682 & 71.04 \\
 $^{16}${O} & 8 & 1.073 & $-0.1731$ & 0.016\,77 & 0.9170 & 127.4 \\
 $^{17}${O} & 8 & 1.073 & $-0.1732$ & 0.016\,77 & 0.9171 & 127.5 \\
 $^{18}${O} & 8 & 1.074 & $-0.1732$ & 0.016\,77 & 0.9172 & 127.5 \\
 $^{19}${F} & 9 & 1.390 & $-0.1995$ & 0.017\,15 & 1.207 & 212.5 \\
 $^{20}${Ne} & 10 & 1.747 & $-0.2260$ & 0.017\,47 & 1.539 & 334.5 \\
 $^{21}${Ne} & 10 & 1.747 & $-0.2260$ & 0.017\,47 & 1.539 & 334.6 \\
 $^{22}${Ne} & 10 & 1.747 & $-0.2260$ & 0.017\,47 & 1.539 & 334.7 \\
 \hline
\end{tabular}
  \caption{The LbL contributions to the Lamb shift of the $1s$ state in a
light two-body muonic atom. The contributions are given in units of
$\alpha^3(Z\alpha)^2m_r$ and meV. The results are given for the total LbL
contribution and for its components (see Fig.~\ref{fig:LbL}). We present
in the table the central values, while the accuracy of the calculation is
discussed in the text.}
  \label{t:1s}
\end{table*}

\begin{table*}
\begin{tabular}{cc|ccccc}
 \hline
 Ion&Z&$\Delta E_{1:3}$&$\Delta E_{2:2}$&$\Delta E_{3:1}$&$\Delta E_{\rm LbL}(1s)$&$\Delta E_{\rm LbL}(1s)$ \\
     &  & [$\alpha^3(Z\alpha)^2m_r$] & [$\alpha^3(Z\alpha)^2m_r$] & [$\alpha^3(Z\alpha)^2m_r$] & [$\alpha^3(Z\alpha)^2m_r$] &[meV]\\
 \hline
 $^{1}${H} & 1 & 0.000\,6323 & $-0.000\,9114$ & 0.000\,6323 & 0.000\,3532 & 0.000\,6941 \\
 $^{2}${H} & 1 & 0.000\,6592 & $-0.000\,9498$ & 0.000\,6592 & 0.000\,3687 & 0.000\,7631 \\
 $^{3}${H} & 1 & 0.000\,6686 & $-0.000\,9632$ & 0.000\,6686 & 0.000\,3740 & 0.000\,7880 \\
 $^{3}${He} & 2 & 0.004\,404 & $-0.003\,188$ & 0.001\,101 & 0.002\,317 & 0.019\,53 \\
 $^{4}${He} & 2 & 0.004\,431 & $-0.003\,207$ & 0.001\,108 & 0.002\,332 & 0.019\,83 \\
 $^{6}${Li} & 3 & 0.013\,19 & $-0.006\,236$ & 0.001\,465 & 0.008\,416 & 0.1625 \\
 $^{7}${Li} & 3 & 0.013\,21 & $-0.006\,246$ & 0.001\,468 & 0.008\,432 & 0.1633 \\
 $^{9}${Be} & 4 & 0.028\,39 & $-0.009\,825$ & 0.001\,774 & 0.020\,34 & 0.7027 \\
 $^{10}${B} & 5 & 0.050\,96 & $-0.013\,83$ & 0.002\,039 & 0.039\,17 & 2.117 \\
 $^{11}${B} & 5 & 0.050\,99 & $-0.013\,84$ & 0.002\,040 & 0.039\,19 & 2.121 \\
 $^{12}${C} & 6 & 0.081\,68 & $-0.018\,21$ & 0.002\,269 & 0.065\,75 & 5.127 \\
 $^{13}${C} & 6 & 0.081\,72 & $-0.018\,21$ & 0.002\,270 & 0.065\,78 & 5.132 \\
 $^{14}${N} & 7 & 0.1210 & $-0.022\,87$ & 0.002\,470 & 0.1006 & 10.69 \\
 $^{15}${N} & 7 & 0.1210 & $-0.022\,88$ & 0.002\,470 & 0.1006 & 10.70 \\
 $^{16}${O} & 8 & 0.1693 & $-0.027\,78$ & 0.002\,645 & 0.1442 & 20.03 \\
 $^{17}${O} & 8 & 0.1693 & $-0.027\,78$ & 0.002\,646 & 0.1442 & 20.04 \\
 $^{18}${O} & 8 & 0.1694 & $-0.027\,79$ & 0.002\,646 & 0.1442 & 20.06 \\
 $^{19}${F} & 9 & 0.2269 & $-0.032\,90$ & 0.002\,801 & 0.1968 & 34.64 \\
 $^{20}${Ne} & 10 & 0.2938 & $-0.038\,19$ & 0.002\,938 & 0.2585 & 56.20 \\
 $^{21}${Ne} & 10 & 0.2938 & $-0.038\,19$ & 0.002\,937 & 0.2586 & 56.23 \\
 $^{22}${Ne} & 10 & 0.2938 & $-0.038\,20$ & 0.002\,938 & 0.2586 & 56.24 \\
 \hline
\end{tabular}
  \caption{The LbL contributions to the Lamb shift of the $2s$ state in a light two-body muonic atom. The contributions are given in units of $\alpha^3(Z\alpha)^2m_r$ and meV. The results are given for the total LbL contribution and for its components (see Fig.~\ref{fig:LbL}). We present in the table the central values, while the accuracy of the calculation is discussed in the text.}
  \label{t:2s}
\end{table*}

\begin{table*}
\begin{tabular}{cc|ccccc}
 \hline
 Ion&Z&$\Delta E_{1:3}$&$\Delta E_{2:2}$&$\Delta E_{3:1}$&$\Delta E_{\rm LbL}(1s)$&$\Delta E_{\rm LbL}(1s)$ \\
     &  & [$\alpha^3(Z\alpha)^2m_r$] & [$\alpha^3(Z\alpha)^2m_r$] & [$\alpha^3(Z\alpha)^2m_r$] & [$\alpha^3(Z\alpha)^2m_r$] &[meV]\\
 \hline
 $^{1}${H} & 1 & 0.000\,1116 & $-0.000\,3265$ & 0.000\,1155 & $-0.000\,095\,43$ & $-0.000\,1875$ \\
 $^{2}${H} & 1 & 0.000\,1300 & $-0.000\,3543$ & 0.000\,1300 & $-0.000\,094\,24$ & $-0.000\,1951$ \\
 $^{3}${H} & 1 & 0.000\,1353 & $-0.000\,3642$ & 0.000\,1353 & $-0.000\,093\,55$ & $-0.000\,1971$ \\
 $^{3}${He} & 2 & 0.002\,065 & $-0.001\,848$ & 0.000\,5161 & 0.000\,7332 & 0.006\,180 \\
 $^{4}${He} & 2 & 0.002\,095 & $-0.001\,867$ & 0.000\,5237 & 0.000\,7518 & 0.006\,394 \\
 $^{6}${Li} & 3 & 0.008\,568 & $-0.004\,338$ & 0.000\,9520 & 0.005\,182 & 0.1001 \\
 $^{7}${Li} & 3 & 0.008\,597 & $-0.004\,349$ & 0.000\,9552 & 0.005\,203 & 0.1007 \\
 $^{9}${Be} & 4 & 0.021\,43 & $-0.007\,548$ & 0.001\,339 & 0.015\,22 & 0.5258 \\
 $^{10}${B} & 5 & 0.041\,72 & $-0.011\,30$ & 0.001\,669 & 0.032\,09 & 1.734 \\
 $^{11}${B} & 5 & 0.041\,76 & $-0.011\,31$ & 0.001\,670 & 0.032\,12 & 1.738 \\
 $^{12}${C} & 6 & 0.070\,29 & $-0.015\,51$ & 0.001\,952 & 0.056\,73 & 4.423 \\
 $^{13}${C} & 6 & 0.070\,33 & $-0.015\,52$ & 0.001\,954 & 0.056\,76 & 4.429 \\
 $^{14}${N} & 7 & 0.1076 & $-0.020\,08$ & 0.002\,196 & 0.089\,72 & 9.535 \\
 $^{15}${N} & 7 & 0.1076 & $-0.020\,08$ & 0.002\,197 & 0.089\,75 & 9.543 \\
 $^{16}${O} & 8 & 0.1540 & $-0.024\,93$ & 0.002\,406 & 0.1315 & 18.27 \\
 $^{17}${O} & 8 & 0.1540 & $-0.024\,94$ & 0.002\,407 & 0.1315 & 18.28 \\
 $^{18}${O} & 8 & 0.1541 & $-0.024\,94$ & 0.002\,408 & 0.1315 & 18.29 \\
 $^{19}${F} & 9 & 0.2098 & $-0.030\,03$ & 0.002\,590 & 0.1824 & 32.10 \\
 $^{20}${Ne} & 10 & 0.2751 & $-0.035\,31$ & 0.002\,751 & 0.2425 & 52.72 \\
 $^{21}${Ne} & 10 & 0.2751 & $-0.035\,32$ & 0.002\,751 & 0.2425 & 52.74 \\
 $^{22}${Ne} & 10 & 0.2751 & $-0.035\,32$ & 0.002\,751 & 0.2426 & 52.76 \\
 \hline
\end{tabular}
  \caption{The LbL contributions to the Lamb shift of the $2p$ state in a light two-body muonic atom. The contributions are given in units of $\alpha^3(Z\alpha)^2m_r$ and meV. The results are given for the total LbL contribution and for its components (see Fig.~\ref{fig:LbL}). We present in the table the central values, while the accuracy of the calculation is discussed in the text. That is a nonrelativistic calculation and therefore the results for $2p_{1/2}$ and $2p_{3/2}$ are the same.}
  \label{t:2p}
\end{table*}

\section{Conclusions}

In conclusion, we have derived a representation for an effective potential
induced by the virtual Delbr\"uck scattering in the leading nonrelativistic
approximation. We have obtained its numerical values in a number of points
in the coordinate space and found an efficient Pad\'e approximation.
The accuracy of the Pad\'e approximation is the highest for $m_er<1$, which
allowed us to find the contributions to the Lamb shift of the low states
in light two-body muonic atoms. We estimate the accuracy of the numerical
evaluation as at the level of one part in a thousand, which is higher than the accuracy of the leading nonrelativistic approximation by itself.

The uncertainty of the Pad\'e approximation for the potential is the best for $m_er<1$ (at the level of $10^{-3}$), and it gradually increases to the
few-percent level for $m_er\simeq10$. The data of the numerical evaluation of the potential itself at higher $m_er$ are not accurate enough; however, the Pad\'e approximation is constrained by the long-distance asymptotic behavior, which we have established by an independent evaluation.

In particular, we have tabulated the related contributions to the Lamb shift of the $1s, 2s, 2p$ states in muonic atoms with $Z\leq 10$. Those states are sufficient for two important problems, namely, for a theory of the $n=2$ Lamb shift
and of the Lyman-$\alpha$ interval.

We have also compared the results for the virtual-Delbr\"uck-scattering
contribution and the Wichmann-Kroll one. At $Z=1$ they are comparable
(being of opposite signs). They increase with the value of $Z$, but
the Wichmann-Kroll one increases faster. At $Z=10$ the virtual-Delbr\"uck-scattering contribution is between 10 and 20\% of the Wichmann-Kroll contribution depending on the state.

\section*{Acknowledgments}

The work was supported in part by RSF (under grant \# 17-12-01036). The work
on calculation of the long-distance behavior was also supported by
DFG (Grant No. KA 4645/1-1). The authors are grateful to Andrzej Czarnecki,
Aleksander Milstein, Akira Ozawa, Krzysztof Pachucki, and Thomas Udem
for useful and stimulating discussions.


\end{document}